\input harvmac
\def\abstract#1{
\vskip .5in\vfil\centerline
{\bf Abstract}\penalty1000
{{\smallskip\ifx\answ\bigans\leftskip 2pc \rightskip 2pc
\else\leftskip 5pc \rightskip 5pc\fi
\noindent\abstractfont \baselineskip=12pt
{#1} \smallskip}}
\penalty-1000}
\def\fc#1#2{{#1\over #2}}
\def\frac#1#2{{#1\over #2}}

\def\br{\hfill\break}
\def\ni{\noindent}

\def\al{\alpha}
\def\eps{\epsilon}

\def\bx#1{{\bf #1}}
\def\cx#1{{\cal #1}}

\def\us#1{\underline{#1}}
\def\hth/#1#2#3#4#5#6#7{{\tt hep-th/#1#2#3#4#5#6#7}}
\def\nup#1({Nucl.\ Phys.\ $\us {B#1}$\ (}
\def\plt#1({Phys.\ Lett.\ $\us  {B#1}$\ (}
\def\cmp#1({Comm.\ Math.\ Phys.\ $\us  {#1}$\ (}
\def\prp#1({Phys.\ Rep.\ $\us  {#1}$\ (}
\def\prl#1({Phys.\ Rev.\ Lett.\ $\us  {#1}$\ (}
\def\prv#1({Phys.\ Rev.\ $\us  {#1}$\ (}
\def\mpl#1({Mod.\ Phys.\ Let.\ $\us  {A#1}$\ (}
\def\atmp#1({Adv.\ Theor.\ Math.\ Phys.\ $\us  {#1}$\ (}
\def\ijmp#1({Int.\ J.\ Mod.\ Phys.\ $\us{A#1}$\ (}

\def\IP{{\bx P}}

\input epsf
\noblackbox
%
%\draftmode
%%%%%%%%%%%%%%%%%%%%%%%%%%%%%%%%%%%%%%%%%%%%%
\def\rk{{\rm rk\, }}

\def\st{\tilde{s}}\def\wh{\hat{W}}
\def\kc#1{K^{(#1)}}
\lref\Rge{A. Klemm, W. Lerche, P. Mayr, C. Vafa, N. Warner,
                \nup {477} (1996) 746;\br
S. Katz, A. Klemm and C. Vafa, \nup {497} (1997) 173;\br
S. Katz, P. Mayr and C. Vafa, \atmp 1 (1998) 53.}{
\lref\Rbm{P. Berglund and P. Mayr, \atmp 2 (1999) 1307.}
%%%%%%%%%%%%%%%%%%%%%%%%%%%%%%%%%%%%%%
\vskip-2cm
\Title{\vbox{
\rightline{\vbox{\baselineskip12pt\hbox{CERN-TH/99-330}
                                  \hbox{hep-th/9910268}}}}}
{Conformal Field Theories on K3}
\vskip -1cm
\centerline{\titlefont and}
\vskip0.2cm
\centerline{\titlefont Three-Dimensional Gauge Theories}

\abstractfont 

%{Title}
\vskip 1cm
\centerline{P. Mayr}
\vskip 0.6cm
\centerline{\it Theory Division, CERN, 1211 Geneva 23, 
Switzerland}
\vskip 0.3cm
\abstract{
According to a recent conjecture, the moduli space of the
heterotic conformal field theory on a $G\subset$ ADE singularity of
an ALE space is equivalent to the moduli space of a pure
$\cx N=4$ supersymmetric three-dimensional gauge theory with 
gauge group $G$. We establish this relation using geometric engineering
of heterotic strings and generalize it to theories with non-trivial matter 
content. A similar equivalence is found between
the moduli of heterotic CFT on isolated
Calabi--Yau 3-fold singularities and two-dimensional Kazama-Suzuki coset 
theories.
}

\Date{\vbox{\hbox{\sl {October 1998}}
}}
\goodbreak

\parskip=4pt plus 15pt minus 1pt
\baselineskip=15pt plus 2pt minus 1pt

\def\ap{{\alpha'}}

%%%%%%%%%%%%%%%%%%%%%%%%%%%%%%%%%%%%%%%%%%%%%%%%%%%%%%%%%%%%%%%
\newsec{Introduction}
%%%%%%%%%%%%%%%%%%%%%%%%%%%%%%%%%%%%%%%%%%%%%%%%%%%%%%%%%%%%%%%
In a recent paper 
\ref\witade{E. Witten, {\it Heterotic string conformal field theory and A-D-E
                  singularities}, hep-th/9909229.}%
, the exact world-sheet instanton
corrected moduli space of the heterotic conformal field theory on
a $G=A_1$ singularity of K3 has been determined in the $M_{str}\to \infty$
limit and found to be given by the Atiyah-Hitchin manifold 
\ref\Rah{M.F. Atiyah and N. Hitchin, {\it The geometry and dynamics of magnetic monopoles}
(Princeton University Press, 1988),}%
. This is the same moduli as 
that of the three-dimensional pure $\cx N=4$ theory 
\ref\swtd{N. Seiberg and E. Witten, {\it 
Gauge dynamics and compactification to three dimensions}, hep-th/9607163.}
with gauge group $G=A_1$. 
It has been conjectured in \witade\ that this relation between 
the moduli spaces of the conformal field theory and  the 3d gauge theories
holds for any $G$. For $G=A_n$, the proposal is verified by the analysis of
\ref\Rsen{A. Sen, \atmp 1 (1998) 115.}.

The three dimensional $\cx N=4$ supersymmetric gauge theories can 
be studied by a circle compactification of the four dimensional
$\cx N=2$ theories \swtd. The instantons in the three dimensional
theory correspond to monopoles and dyons in four dimensions.
This situation can also be studied in the field theory limit 
of a type II compactification on a Calabi--Yau $W$ times $S^1$.
In this theory, the charged states of the field theory are represented 
by D-brane wrappings on small $p$-cycles $C_p$ of a local singularity in $W$;
in the limit $\ap\to 0$ with $V(C_p)/(\ap)^{p/2}$ fixed, most
of the fundamental states and gravity decouple and one is left with
the light spectrum of the field theory. 
{\it E.g.}, in the type IIB theory,
the monopole in four dimensions is a D3 brane wrapped on a small
3-cycle $C_3$ of a singularity in a local patch of $W$. The instanton in 
3d is the Euclidean wrapping of the D3 brane on $C_3\times S^1$. 

The gauge coupling $g_3$ of the three-dimensional field 
gets perturbative contributions at tree level and one-loop
as well as possible instanton corrections. The instanton
correction is of the form $e^{-(I+i\sigma)}$, where $\sigma$
is the scalar dual to the photon and the instanton action
$I$ is approximately $2\pi M r$ for large $r$, where $M$ is 
the mass of the monopole in four dimensions. In the type II 
theory the latter is given by the volume of the brane 
wrapping $V(C_3)$. 

The three dimensional gauge theories enjoy often interesting 
dualities that equate the Coulomb branch of a certain theory
with the Higgs branch of another 
\ref\Ris{K. Intriligator and N. Seiberg, \plt 387 (1996) 513.}%
\ref\Rtdd{J. de Boer, K. Hori, H. Ooguri and Y. Oz, \nup 493 (1997) 101;\br
M. Porrati and A. Zaffaroni, \nup 490 (1997) 107.}%
\ref\Rhov{K. Hori, H. Ooguri and C. Vafa, \nup 504 (1997) 147.}.
In the type II string,
this duality is interpreted as a T-duality on the $S^1$ which 
takes the type IIB on $W\times S^1(r)$ to 
type IIA on $W\times S^1(1/r)$. If $W$ is a singularity of 
small 3-cycles as in the previous discussion, the field theory 
limit of type IIA describes a theory of hypermultiplets parametrizing
the volumes of, and RR fields on, the cycles $C^i_3$. 
This is the theory that should be naturally
identified with the heterotic CFT in the duality discussed in this
paper. The D3 instantons of the type IIB theory become Euclidean 
D2 branes wrapped on $C_3$ after the T-duality.

In the heterotic CFT, the expansion is in terms of world-sheet instantons
and the corresponding instanton action of the form $e^{-\fc{V(c)}{\ap}}$,
where $c$ is a 2-cycle in the K3 of the heterotic compactification
on which a Euclidean fundamental string is wrapped. 
Moreover tree-level and one-loop of the field theory correspond 
to the $\ap^0$ and $\ap^1$ contribution to the heterotic string metric, respectively
\witade.
To obtain the Hyper-K\"ahler moduli space of the field theory
one takes a similar field theory limit
of the heterotic string with $\ap\to 0$ at fixed  $ V(c)/\ap$. 
The moduli space of hypermultiplets describes
the local deformations of an ADE singularity of an ALE space that
governs the local patch of the K3 in this limit. 

Although the expansion of the gauge theory and the CFT is morally speaking
around the same point, namely in the limit of large instanton action
or large volume of the 2-cycles $c$ in the heterotic string or
$C$ in the type IIA theory, respectively, it would be hasty to 
conclude that the corresponding instanton expansions are exactly the
same. In fact it would be surprising if the wrappings of the 
D2 brane instantons of the type IIA theory would follow the same 
rule as that of the fundamental world-sheet instantons of the 
heterotic CFT. It would be interesting to relate the two instanton
expansions.

In this note we derive the conjecture of ref.\witade\ by using
heterotic/type II duality to show that the type II geometry $W$
associated to the heterotic string the ADE singularity is
the appropriate, known Calabi--Yau singularity of small 3-cycles 
that yields the $G$ gauge theory upon type IIB compactification 
on $W\times S^1$. We will also consider various generalizations of the conjecture.

\newsec{Field theory limit}
Consider the heterotic string compactification on $K3\times T^2$, 
where the K3 will be replaced by the ADE singularity momentarily. 
In general this is dual to type IIA on a K3 fibered Calabi--Yau
3-fold $W$. If we assume that the bundle factorizes on K3$\times T^2$
and the heterotic K3 is elliptically fibered, the four-dimensional duality
can be pushed up to six dimensions, between F-theory on the,
now elliptically fibered, Calabi--Yau $W$ and heterotic string on K3. 

Let us first show that the 
the field theory limit of the heterotic string described by the
CFT on the ADE singularity with a Hyper-K\"ahler moduli space
corresponds to a - special - field theory limit of the type IIA theory 
The heterotic parameters behave as
\eqn\hetlim{
g_{het}\to 0,\qquad \ap_{het}\to 0,\qquad V(c)/\ap_{het}={\rm const.},
}
where $c$ stand collectively for a 2-cycle in the ADE singularity.

First note that all $p$-cycles of K3 except for the cycles $c$ of the
singularity grow as $\ap^{-p/2}$ in the limit \hetlim. This holds
in particular for the two universal classes $E$ and $B$ of the elliptically
fibered K3 with a section, where $E$ is the class of the fiber and
$B$ the class of a section. Consider the singularity from a 
collision of singular elliptic fibers at a point $s_0$ of $B$.
The transverse dimensions to the singularity are described by 
$E$ and $B$, which become non-compact in the above limit.
Therefore we are left with the local singularity of a non-compact ALE space.

The four-dimensional compactification above can be considered as 
a compactification of the six-dimensional duality between type IIA
on K3 and heterotic string on $T^4$ on a base $\IP^1$ denoted by $B$. 
From the relations in 
\ref\witcom{E. Witten, \nup 443 (1995) 85.}
one can then easily see that the four-dimensional
coupling constants are related by
\eqn\fdcoup{
\fc{1}{g_{4,het}^2}\sim\fc{V_{het}(B)}{g_{6,het}^2}\sim  V_{II}(B),\qquad 
\fc{1}{g_{4,II}^2} \sim\fc{V_{II}(B)}{g_{6,II}^2}\sim V_{het}(B).}
Note that $B$ is the base of the elliptically fibered K3 of
the heterotic theory and the K3 fibration of $W$ in the 
type IIA theory, respectively.
Since $V_{het}(B)$ becomes large, we have a weakly coupled type IIA theory.
To have a weakly coupled heterotic string, $V_{II}(B)$ becomes large, too.

Moreover the type IIA string is obtained from wrapping a 
heterotic 5-brane on $T^2$ times the elliptic fiber of K3.
For generic volume of $T^2$ this 4-cycle becomes very large 
and therefore $\ap\to 0$ in the type II theory. However
as we will show in the next section, the limit we have to take is not
the naive $\ap\to 0$ limit: to the small 2-cycles $c_i$ in K3 correspond small 
3-cycles $C_i$ in $W$ with a fixed volume $V(C_i)/\ap_{II}^{3/2}$.

Above 
we have recovered the 
definition of the field theory limit of the dual type II string
studied in the geometric engineering approach of type II strings \Rge\ on 
Calabi--Yau singularities. The fact that it is the fiber of 
the heterotic K3 $Z$ that gets
large (and not just the 4-cycle of the  5-brane wrapping) implies a special
limit in the complex structure of the Calabi--Yau singularity.
We will show in the next section that the special singularities
one obtains in this way are precisely those used in \Rbm\ to extend 
the geometric engineering approach of \Rge\ to describe heterotic moduli 
spaces by F-theory limits of Calabi--Yau singularities.
The subsequent sections may be understood without going through 
the mathematics of the next section.

\newsec{Dual type IIA singularities and stable degenerations}
A second consequence of the large elliptic fiber of $Z$ is a
limit in the complex structure of $W$  that is known as a stable degeneration
\ref\Rmv{D.R. Morrison and C. Vafa, \nup 476 (1996) 437; \nup 473 (1996) 74.}%
\ref\Rfmw{R. Friedman, J. Morgan and E. Witten, \cmp 187 (1997) 679.}%
\ref\Ram{P.S. Aspinwall and D.R. Morrison, \nup 503 (1997) 533.}.
In particular it was proposed  in \Rmv\ that if $W$ is a K3 manifold, 
one can see the heterotic $T^2$ in $W$ on the nose in 
the limit of the stable degeneration,  which in heterotic terms  corresponds to 
a very large $T^2$.
Since we are interested in the complex structure of $W$ parametrized by the
hypermultiplets of the type IIA theory, we can consider F-theory compactification on $W$
dual to heterotic on $Z$ without the extra $T^2$. The 
type IIA theory is related to F-theory by the 
extra $T^2$ compactification which will
affect only the K\"ahler moduli space of the Calabi--Yau $W$.

In the following we will describe in some detail general deformations of the
stable degeneration and make contact with the field theory limit of the type IIA on 
non-compact singularities considered in \Rbm. We will verify the identifications 
made in \Rmv\ and add also the information about the bundles on the heterotic manifold $Z$ 
in this way. This will serve as a starting point for the identification of 
the type II singularity dual to the ADE singularity of the heterotic string in the
next section.

The elliptically fibered manifold $W$ for F-theory compactification in
Weierstrass form is given by
\eqn\ftmf{
p_W=y^2+x^3+xf+g=0,}
with $f$ and $g$ some functions depending on the coordinates on the base $B$ of the elliptic
fibration $\pi:W\to B$. To describe the limit in complex structure we are interested in, we 
consider families of the F-theory manifold $W$ in a fibration over the complex plane, 
similarly as in \Ram. Let us write $\ftmf$ in more detail as 
\def\be{\beta}
\eqn\ftmfii{
p_W=y^2+x^3+x\sum_\al s^{4-\al}\st^{4+\al}f^{(4)}_{\al}+\sum_\al s^{6-\al}\st^{6+\al}f^{(6)}_{\al},}
where $f^{(k)}_{\al}$ are constants if $W$ is a K3, but more generally functions on the 
$n-1$ dimensional base $B'$ of the K3 fibration $\pi':\ W\to B'$ of 
the Calabi--Yau $n+1$-fold W.
Moreover $s$ and $\st$ are homogeneous coordinates on the base $\IP^1$ of the elliptic
fibration of the  K3 fiber.

To perform the limit, we consider families $Y$ of $W$ 
in a fibration over another $\IP^1$ parametrized by $\mu$
$$
p_Y=y^2+x^3+x\sum_{\al,\be} s^{4-\al}\st^{4+\al}\mu^{4-\be}f^{(4)}_{\al} a^{(4)}_{\al,\be}
+\sum_{\al,\be} s^{6-\al}\st^{6+\al}\mu^{6-\be}f^{(6)}_{\al} a^{(6)}_{\al,\be} ,
$$
where $a_{\al,\be}^{(k)}$ are some complex constants that we will set to 0 or 1 in the
following.
In general, the fiber of $Y$ at $\mu=0$ is a (certain deformation) of the original manifold
$W$. However if we restrict the complex structure of $Y$ such that $a^{(k)}_{\al,\be}=0$ for 
$\al+\be>k$ the fiber at $\mu=0$ becomes 
$$
p_Y=y^2+x^3+xs^{4}\sum_{\al\leq0} s^{-\al}\st^{4+\al}f^{(4)}_{\al} 
+s^{6}\sum_{\al\leq0} s^{-\al}\st^{6+\al}f^{(6)}_{\al} ,
$$
which has a non-minimal singularity at $y=x=s=0$. After a  blow up of $Y$,
$$
y=\rho^3y,\ x=\rho^2x,\ s=\rho s,\ \mu = \rho \mu,
$$
we arrive at 
$$
p_{Y'}=y^2+x^3+x\sum_{\al,\be} s^{4-\al}\st^{4+\al}\mu^{4-\be}\rho^{4-\al-\be}f^{(4)}_{\al} 
+\sum_{\al,\be} s^{6-\al}\st^{6+\al}\mu^{6-\be}\rho^{6-\al-\be}f^{(6)}_{\al}.
$$
The singular fiber at $\mu =0 $ is now replaced by a reducible fiber $\wh$ with two components
corresponding to $\mu=0$ and $\rho=0$:
\eqn\twocomp{\eqalign{
p_{W_1}=p_0+p_+=&(y^2+x^3+x f^{(4)}_0 +f^{(6)}_0 )+\cr
&(x\sum_{\al<0} \st^{4+\al}\rho^{-\al}f^{(4)}_{\al} 
+\sum_{\al<0} \st^{6+\al}\rho^{-\al}f^{(6)}_{\al} ), \cr
p_{W_2}=p_0+p_-=&(y^2+x^3+x f^{(4)}_0 +f^{(6)}_0 )+\cr
&(x\sum_{\al>0} s^{4-\al}\mu^{\al}f^{(4)}_{\al}
+\sum_{\al>0} s^{6-\al}\mu^{\al}f^{(6)}_{\al} ) .\cr}
}
The two components $W_1$ and $W_2$ 
intersect over the locus $\mu=\rho=0$ described by $p_0=0$.
According to \Rmv\ this intersection is to be identified with the heterotic manifold $Z$.

This proposal can be made precise using local mirror symmetry 
which allows to also extract the heterotic bundle \Rbm.
As shown there, a type IIA compactification on certain non-compact Calabi--Yau
singularities $\cx W$ results in a gravity free field theory with the moduli space
of holomorphic stable bundles on a compact Calabi--Yau $Z_n$ defined as a certain 
hypersurface in $\cx W$. For appropriate structure group of the bundle this data
can be interpreted as a heterotic compactification in the point particle limit.
From comparison with \twocomp, one observes that 
if we consider separately a manifold $\cx W_1$ described by $p_{W_1}=0$, it describes precisely 
a non-compact Calabi--Yau singularity of the type considered in \Rbm. 
The singularity $\cx W_1$ indeed describes the moduli space of holomorphic stable bundles
$V$ on $Z:\ p_0=0$.
Moreover we can use the identifications for the
singularity $\cx W_1$ derived in \Rbm\ to read off the moduli of the bundle $V$ on $Z$.

\newsec{Chains and cycles in $\wh$}
Let us show now how the vanishing 2 cycles $c_i$ 
of the heterotic ADE singularity correspond to shrinking 3-cycles in the dual 
type II manifold $\wh$. The latter  
support the D2 instantons of the field theory limit of the type IIA theory 
and the monopole wrappings of the T-dual type IIB theory that represents the
Coulomb branch of the three-dimensional gauge theory.

Let us start with the illustrative and simpler 
case of F-theory on a K3 $W$ dual to heterotic string on $T^2$.
It shares many of the relevant aspects with the case of the heterotic string
on K3 that will
be obtained by fibering over an extra  $\IP^1$.

Before the stable degeneration, $W$ is an elliptic fibration over a $\IP^1$ with 
24 singular elliptic fibers located at the zeros of the discriminant $\Delta=4f^3+27g^2$,
where $f$ and $g$ are degree 8 and 12 polynomials in the variable $s$ on the base $\IP^1$,
respectively. After the stable degeneration, $W$ has split into the two 
components $W_1$ and $W_2$ described by eq. \twocomp, intersecting over the
elliptic curve $E:\ p_0=0$. The base of the elliptic fibrations of the elliptic 
rational surfaces $W_i$
are two $\IP^1$'s that intersect over a point, $\mu=\rho=0$.
There are 12 points on each $\IP^1$ above which the elliptic fibers
becomes singular:

\vskip 0.5cm
{\baselineskip=12pt \sl
\goodbreak\midinsert
\centerline{\epsfxsize 3.6truein\epsfbox{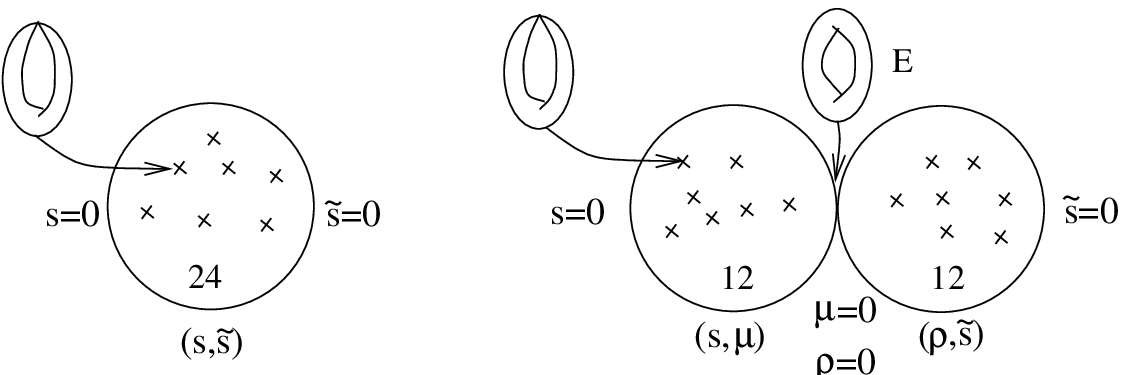}}
\leftskip 1pc\rightskip 1pc \vskip0.3cm
\noindent{\ninepoint  \baselineskip=8pt 
{{\bf Fig. 1:}
The original K3 and the generic stable deformation.}
}\endinsert}\vskip -0.4cm

There are two basic degenerations of this elliptic fibration. First we can bring together
singular elliptic fibers of a single component, say  $W_1$. This introduces a singularity of
small 2-cycles in $W_1$ with a type determined by the classification of Kodaira.
Since the relative location of the singular elliptic fibers is described
by $p_+$ in eq.\twocomp, with parameters that translates to that of the heterotic bundle as described in \Rbm,
this corresponds to a degeneration of the gauge bundle in the heterotic string on $T^2$. 

The second kind of degeneration consists of bringing together a singular fiber from 
$W_1$ with another singular fiber from $W_2$. This is only possible at the
intersection of the two base $\IP^1$'s at $\mu=\rho=0$. Again we obtain a shrinking 2-cycle $c$
in $\wh$. However note that $c$ intersects $E$ along a 1-cycle $\gamma$. 

To see the cycle $c$ and the fact that its volume vanishes if and only if the 
1-cycle $\gamma$ shrinks, consider the local situation in $W_1$ described by choosing
\eqn\Elocl{
f=f_0+\rho f_1=(-3a^2)+\rho f_1,\qquad g=g_0+\rho g_1=(2a^3-\eps)+\rho g_1,}
in \ftmf.
For small $\eps$, the heterotic torus at $\rho=0$ has a small $S^1$ described by
$$
\gamma:\ y^2+x^2=\eps.
$$
Away from $\rho=0$ there is still an $S^1$, $\gamma(\rho)$,
with a radius $\sqrt{\eps(\rho)}$ depending on the 
value of $\rho$. From the vanishing of the discriminant $\Delta=\Delta_0+\rho \Delta_1$
we see that $\gamma(\rho)$ shrinks to zero size 
at a point $\rho_0=-\Delta_0/\Delta_1$. We can therefore define
a two 2-chain $\kc1$ in $W_1$ from gluing together 1-cycles $\gamma(\rho)$ along the interval
$\rho=[0,\rho_0]$. 
\vskip 0.5cm
{\baselineskip=12pt \sl
\goodbreak\midinsert
\centerline{\epsfxsize 1.5truein\epsfbox{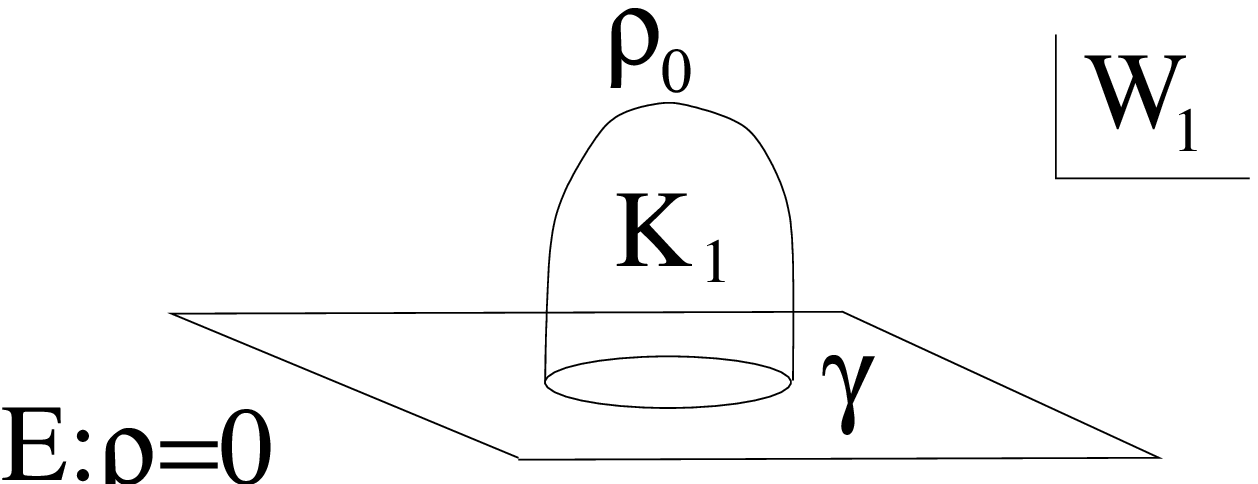}}
\leftskip 1pc\rightskip 1pc \vskip0.3cm
\noindent{\ninepoint  \baselineskip=8pt 
{{\bf Fig. 2:}
The 2-chain $\kc1$ in $W_1$.}
}\endinsert}\vskip -0.4cm

Note that the cycle $\gamma=\gamma(0)$ in $E$ shrinks iff $\Delta_0=0$, which 
implies $\rho_0\to 0$. Thus precisely if $\gamma$ shrinks, $\kc1$ shrinks, too.
In fact the volume of the 2-chain $\kc1$ is determined by the integral 
\eqn\Evolint{
\int_{\kc1} \Omega \sim  \int_{\kc1} \fc{dx\ d\rho}{y} = \fc{\pi \eps}{\sqrt{3a}\ (af_1+g_1)},}
where the proportionality constant depends on the normalization of the 2-form $\Omega$.
In the total manifold $\wh$ we obtain a vanishing 2-cycle from the two vanishing 
2 chains $\kc \al \subset W_\al$ joining along the vanishing cycle $\gamma$. Note that
we can write the discriminant of $\wh$ as $\Delta(\wh)=\Delta(W_1)\cdot \Delta(W_2)$,
which vanishes to second order at $\mu=\rho=0$ due to the first order vanishing
of $\Delta(W_i)$.

Thus this second kind of degeneration is related to a geometric degeneration
of the heterotic torus $E$. This does not make too much sense in the case of the
torus, since its degenerations are very limited and in fact at infinite distance
in the moduli space\foot{However note that in this limit we have $U=\infty=T$ so
we can interpreted the $SU(2)$ gauge symmetry of the type IIA string on 
the vanishing 2-cycle as the heterotic gauge symmetry enhancement at $T=U$ in
the limit $T\to \infty$.}.
We can also combine the two kind of degenerations and bring together several
singular fibers from $W_1$ and $W_2$ at the intersection. From the above this should
correspond to a degeneration of the bundle on top of the singular heterotic geometry.

Having discussed 
this toy example, let us consider the case of heterotic string on a non-trivial
Calabi--Yau $Z_n,\ n>1$, in particular for $Z=K3$. The new ingredient is that 
we can now have geometric singularities of $Z$ at finite distance in the moduli,
in particular ADE singularities of ALE spaces. Let us fiber the 
eight-dimensional picture in Fig. 1 over an extra $\IP^1$, parametrized by $t$.
The local structure of the 3-fold geometry $\wh$ for small 2-cycles $c_i$ in $Z$ is similar
as in eq.\Elocl\ with an extra dependence on the base $\IP^1$:
\eqn\Elocii{
f=f_0(t)+\rho\ f_1(t), \qquad g=g_0(t)+\rho \ g_1(t).}
Similarly we can split the discriminant of the elliptic fibration as 
$\Delta=\Delta_0(t)+\rho\ \Delta_1(t)$. Near a $G$ singularity of the K3,
the zeros of $\Delta_0$ are described by
\eqn\Edis{
\Delta_0=\prod_i^\delta (t-t_i(a_k))=t^\delta+\dots,}
where $a_k,\ k=1,\dots,{\rm rk} \ G$ denote the deformations of the singularity
and $\delta$ is a certain integer that counts 
the number of simple singular fibers
needed to produce the $G$ singularity in the elliptic fibration\foot{{\it E.g.}
$\delta=N$ for $SU(N)$ and more generally $\delta={\rm rk} G +1 +\nu$, where $\nu$ is
the number of trivalent vertices in the Dynkin diagram of $G$.
See ref.\ref\RBal{M. Bershadsky et al., \nup 481 (1996) 215.}.}. 
Classically, the $G$ singularity
is recovered in the limit, where the perturbations $a_k$ become zero.

Moving away from $\rho=0$ to describe the 3-fold $\wh$, the zeros $t_i$ of the discriminant 
$\Delta$ depend on $\rho$:
\eqn\disii{
\Delta=\Delta_0+\rho\ \Delta_1 =\prod_i^\delta (t-t_i(a_k,\rho)).}
Note that for each pair of $t_i(a_k,\rho)$, we can define a 2-cycle $c_{ij}(\rho)$ 
by stringing together 1-cycles in $E$ along the interval $[t_i(a_k,\rho),t_j(a_k,\rho)]$,
very similar as we did in the construction of the 2-chains $\kc \al$. Moreover
the 2-cycle $c_{ij}(\rho)$ attaches to a 2-cycle $c_{ij}$ of the K3 $Z$ for $\rho=0$.
In addition it shrinks to zero size at a point $\rho=\rho_{ij}$ where
\eqn\Evtc{t_i(a_k,\rho_{ij})=t_j(a_k,\rho_{ij}).}
Very similar as in the previous case we can now build 3-chains $\kc \al _{ij}$ by gluing
together the 2-spheres $c_{ij}(\rho)$ along the interval $[0,\rho_{ij}]$. 

Finally note that for very similar reasons as for the 2-chains constructed
before, the volume of the 3-chain $\kc \al _{ij}$ vanishes precisely if the 
2-cycle $c_{ij}$ shrinks and in particular all 3-chains shrink near the singularity of
the heterotic K3 $Z$
\eqn\Edegi{
a_k \to 0 \ \Rightarrow \ \rho_{ij}\to 0.}
One can calculate also the volume of the individual
3-chains $\kc \al _{ij}$ similarly as in \Evolint.

We have therefore succeeded to establish a 1-1 correspondence between the small
2-cycles $c_{ij}$ of the resolved ADE singularity and small 3-cycles $C_{ij}=\kc 1 _{ij} \cup 
\kc 2 _{ij}$ of a singularity reached in the complex structure
of the type II compactification manifold $\wh$. The lattice
$H_3(\wh)$ inherits the ADE structure from the lattice $H_2(Z)$. Note that the cycles $C_{ij}$
intersect each other according to the ADE Dynkin diagram. This fits with the physics expectations
that the instanton corrections to the three-dimensional gauge theory represented by
type IIB on $\wh$ arise from the monopoles and dyons, as opposed to the wrappings of 
purely electric cycles.

The singularity of the type II manifold $\wh$ is precisely the one which is known to
produce the $G$ gauge theory in three dimensions in a type IIB compactification
on $\wh\times S^1$ \Rge. Note that we needed the extra $S^1$ to translate from the heterotic
type IIA dual that describes the "Higgs branch" of hypermultiplets, to the Coulomb
branch of the $G$ gauge theory of the type IIB field theory limit, which corresponds to 
the dual theory in the sense of \Ris,\Rtdd. Let us also identify the states dual to 
the monopoles from D3 branes wrapped on $C_{ij}$. In the T-dual type IIA theory they become
D4 branes wrapped on $C_{ij}\times S^1$. In the heterotic theory they become 
5-branes wrapped on $c_{ij}\times T^3$.

Let us consider now possible generalizations of the above picture by moving additional
singular elliptic fibers into the heterotic singularity. As discussed in the eight-dimensional
picture, if we let collide fibers at some point $\rho=\rho'$ of the base of 
the elliptic surface $W_i$, 
this corresponds to a degeneration of the heterotic bundle. Moving this point into
the ADE singularity, $\rho'\to 0$, we expect that in general the gauge bundle is not 
trivial on the singularity. This would correspond to theories with non-perturbative 
heterotic dynamics, such as the extra 
gauge symmetry enhancement discussed in \Rbm. 

Let us first consider a simpler case, where it is easier to find the answer.
In the six-dimensional compactification we can also consider moving in 
additional singular elliptic fibers into the singularity using the $t$ direction.
In this way we can keep the singularity of the elliptic surface
$W_i$, and thus the bundle, fixed
on the singularity. In particular, if we
choose a totally trivial gauge bundle there are 24 small instantons located at points 
$t=t_i$ of the base of Z. The case where some of these positions coincide with the 
location of the ADE singularity has been studied in great detail in 
\ref\Rbi{K. Intriligator, \nup 496 (1997) 177;\br
J. Blum and K. Intriligator, \nup 506 (1997) 199, \nup 506 (1997) 223.}%
\Ram.
With some effort one can show that after moving in such points there are new 3-cycles
in $\wh$ that correspond to adding matter in the gauge theory following the rules
of geometric engineering of type II strings. Note that we can only get matter
that can be obtained from adjoint breaking of a higher rank gauge group 
\ref\Rkv{S. Katz and C. Vafa, \nup 497 (1997) 146.} in this way.
We will take a shortcut here that uses the results on three-dimensional mirror
symmetry described in \Rtdd\Rhov, which will lead us also to a natural conjecture for new
three-dimensional dualities. 
Specifically, ref.\Rhov\ established a series of three-dimensional dualities using 
geometric engineering of type II on geometric singularities, which is precisely
the situation that we consider. It was shown that if the type IIB
theory on a Calabi--Yau $W$ describes a $U(N)$ theory with $N_f=k$ fundamentals,
than the type IIA theory on the same manifold describes the Higgs branch of a dual
theory with gauge group $U(1)\cdot U(2) \cdot\cdot\cdot U(N-1)\cdot U(N)^{k-2N+1} 
\cdot U(N-1) \cdot\cdot\cdot U(2) \cdot U(1)$. On the other hand \Rbi\Ram\ this is precisely
the gauge theory of $k$ small instantons on a $SU(N)$ singularity of K3!
Thus we see that by considering the field theory limit 
of the heterotic theory with small instantons on the K3 singularity 
it essentially follows from the results in \Rhov\ that the 
heterotic moduli of the $A_{N-1}$ singularity with $k$ instantons is the same 
as that of the three-dimensional $SU(N)$ theory with $N_f=k$.\foot{It is 
worth noticing that this result is consistent with a proposal for
mirror symmetry of F-theory formulated in
\ref\Rpr{E. Perevalov and G. Rajesh, \prl 79 (1997) 2931.}
and derived and
refined in \Rbm. In particular it states that if F-theory on a 3-fold $\tilde W$ describes
the Coulomb branch of $k$ small heterotic instantons on a $G$ singularity 
of K3, than the mirror $\tilde{W}^*$ 
is dual to a heterotic theory on the Higgs branch describing a large 
$G$ instanton with instanton number $k$. After a $T^3$ compactification, the two
theories are related by the three-dimensional duality as above.}

It is natural to wonder whether this relation is true more generally, also
in those cases, where three-dimensional dualities have not yet been established
or even formulated. The prediction would be that the theory dual to the three-dimensional
gauge theory with  gauge group $G$ and $k$ matter multiplets is the 
gauge theory with gauge group $\tilde G$, where $\tilde G$ 
is the extra local gauge group that appears
if $k$ small heterotic instantons collide with a $G$ singularity of K3.

Let us now come back to the case where we move singular fibers of 
the elliptic surface $W_1$ into the 
ADE singularity. We have 
in the moment no good understanding of what the three-dimensional theories associated
to such a degeneration may be, however there is also a plausible candidate for this case
from the geometric engineering of type II strings. Note first that
if we take the base to be $\IP^1\times \IP^1$, we can interpret any of the
two $\IP^1$'s as the base of the heterotic K3. Accordingly we have two heterotic theories
that are related by heterotic-heterotic duality 
\ref\Rhetd{M.J. Duff, R. Minasian and E. Witten, \nup 465 (1996) 413;\br
D.R. Morrison and C. Vafa, \nup 476 (1996) 437; \nup 473 (1996) 74.}. 
An extension of this duality 
for singular K3's and bundle singularities exists and has been described in \Rbm. 
In the previous case we have considered having a $G$ singularity in the K3 with base
$\IP^1$ parametrized by $t$. Of course nothing changes if we consider instead 
only a singularity $G'$ in the heterotic theory with base the other $\IP^1(s,\st)$.
If we now consider a manifold where the $G$ and $G'$ singularities collide,
we have in some sense a collision of two gauge theories with groups $G$ and $G'$. 
This reminds very much of the structure of the conformal field theories engineered from
type II strings in \ref\KMV{S. Katz, P. Mayr and C. Vafa, \atmp 1 (1998) 53.}. 
In this case one has a gauge group $G=\prod SU(k\ n_i)$, with $n_i$ the Dynkin indices
of a gauge group $G'$ and bi-fundamentals in each group factors connected by a link
in the Dynkin diagram. The coupling constants of the gauge theory are in fact related to
the Coulomb parameter of a $G'$ gauge theory which is however broken to 
a trivial $U(1)^{\rk G'}$ factor 
in the region where we observe the 
$G$ gauge theory. However it is also possible
to to move towards the origin of the Coulomb branch of $G$ and $G'$ at the 
same time. One obtains in this limit
an exotic gravity free theory with string like degrees of freedom \KMV. The compactification
of this theory on $S^1$ seems to be a good 
candidate for the three-dimensional theories related to
a collision of a singularity in $W_1$ with an ADE singularity in $Z$.

\newsec{Heterotic CFT on Calabi--Yau 3-fold singularities and
Kazama-Suzuki models in two dimensions}
By the same arguments as in the previous section we can construct
more generally 
a correspondence between $m+1$-chains in the stable degeneration 
of a Calabi-Yau $m+1$-fold $W_{m+1}$ and $m$ cycles in the intersection
of its two components which is an elliptic 
$m$-dimensional Calabi--Yau manifold.
For the case $m=3$ we obtain a correspondence between the heterotic
CFT on the isolated singularity 
\eqn\hstf{
y^2+H(x,s,t)=0,}
and type II strings on a Calabi--Yau 4-fold with a corresponding
singularity of one dimension higher described locally by a similar equation
with $H$ replaced by
$H+w^2$. Here $H=0$ describes the ADE surface singularities
\eqn\adesingst{
\vbox{\offinterlineskip\tabskip=0pt\halign{\strut
$#$~\hfil&\ \ \ $#$,~\hfil\qquad \ \ \ \ \ &$#$~\hfil&\ \ \ $#$~\hfil\cr
H=s^n+t^2+x^2&A_{n-1}&H=s^3+st^3+x^2&E_7,\cr
H=s^n+st^2+x^2&D_{n+1}&H=s^3+t^5+x^2&E_8.\cr
H=s^3+t^4+x^2&E_6&&\cr
}}}
The field theory limit of type IIA theory on these singularities
has been identified in
\ref\Rgvw{S. Gukov, C. Vafa and E. Witten, {\it 
CFT's from Calabi-Yau four-folds}, hep-th/9906070.}
as a Kazama-Suzuki model based on a coset that depends on 
the choice of a RR background on 4-cycles.
The RR fields of type IIA map to gauge fields of the hetrotic string.
We conclude that the heterotic CFT on the 3-fold singularity 
with appropriate gauge field background is on 
the same moduli as the corresponding Kazama-Suzuki model.
A simple consistency check on this equivalence is the following one.
Using linear sigma models it was argued in \witade\ that the
heterotic CFT should be smooth on a singular Calabi--Yau $n$-fold
for any $n$ as long as the gauge fields are trivial on the singularity.
For trivial RR-fields one obtains a massive Kazama-Suzuki model
with no light degrees of freedom. This matches the non-singular 
behavior of the heterotic theory advocated
in \witade. It would be interesting to compare in detail 
the moduli spaces for the more interesting case with non-trivial
background fields, in particular when the heterotic string 
is compactified on the $(2,2)$ theory.

\newsec{Discussion}
The equivalence of the moduli space of the heterotic CFT on ADE singularities and
the moduli space of three dimensional gauge theories, or the dual monopole moduli spaces,
is an interesting tool to study several aspects. {\it E.g.}, the metric of the 
Atiyah-Hitchin manifold is known \Rah\ and one can extract the world-sheet instanton
expansion of the heterotic CFT, similarly as one obtains the world-sheet instantons
of the type II string from mirror symmetry. By an appropriate reparametrization, 
we expect to reproduce the D2 instanton expansion of the type IIA string on $\wh$. 
This allows to analyze a new class of brane wrappings.
It is interesting to note in this context that it is easy to verify that 
the Atiyah-Hitchin metric is compatible with the 
claim in \swtd\ that in the gauge theory variables there is only a single 
instanton contribution; this is a first non-trivial compatibility check. 
Reversing the logic it would also be very interesting to study metrics on 
monopole moduli spaces from the CFT approach. Similarly it would be interesting
to study the relation between instanton corrections to the metric of 
the heterotic 3-fold and the two-dimensional Kazama-Suzuki models. 
\vskip 0.2cm

\ni{\bf Note added:} While preparing this note for publication, two related papers
appeared on the subject. 
Ref.\ref\Rroz{M. Rozali, {\it Hypermultiplet moduli space and three dimensional gauge
                  theories}, hep-th/9910238.}
gives an independent derivation of the conjecture of ref.\witade\ using
M-theory. Ref.%
\ref\Rples{R.S. Aspinwall and R. Plesser, 
{\it Heterotic string corrections from the dual type II string}, hep-th 9910248.}
treats the special case $G=A_1$ case, already analyzed in \witade, using
a "new" variant of geometric engineering for
heterotic strings. In fact this is the same geometric engineering used in this paper\foot{
In particular the definition of the geometric limit on p.6
in \Rples\ is a special example of the
analysis in App. A of \Rbm.} and derived and studied in detail by local mirror symmetry in 
\Rbm\KMV%
\ref\Rge{For an introduction and background material see 
P. Mayr, hep-th 9904115; hep-th/9910216; 
{\it D-branes and F-theory}, lectures the Trieste Spring School 1999, to appear in the
proceedings.}. 

\vskip 0.5cm

\ni
{\bf Acknowledgments:}\br
We would like to thank W. Lerche and Y. Oz for valuable discussions.

\listrefs
\end